\documentclass[twocolumn,english,pra,showpacs,aps]{revtex4-1}

\usepackage[T1]{fontenc}
\usepackage[latin9]{inputenc}
\usepackage{amsmath}
\usepackage{amssymb}
\usepackage{graphicx}
\usepackage{times}
\usepackage{helvet}
\usepackage[normalem]{ulem}

\makeatletter
 
 \@ifundefined{textcolor}{}
 {%
   \definecolor{BLACK}{gray}{0}
   \definecolor{WHITE}{gray}{1}
   \definecolor{RED}{rgb}{1,0,0}
   \definecolor{GREEN}{rgb}{0,1,0}
   \definecolor{BLUE}{rgb}{0,0,1}
   \definecolor{CYAN}{cmyk}{1,0,0,0}
   \definecolor{MAGENTA}{cmyk}{0,1,0,0}
   \definecolor{YELLOW}{cmyk}{0,0,1,0}
 }

\@ifundefined{textcolor}{}{%
 \definecolor{BLACK}{gray}{0}
 \definecolor{WHITE}{gray}{1}
 \definecolor{RED}{rgb}{1,0,0}
 \definecolor{GREEN}{rgb}{0,1,0}
 \definecolor{BLUE}{rgb}{0,0,1}
 \definecolor{CYAN}{cmyk}{1,0,0,0}
 \definecolor{MAGENTA}{cmyk}{0,1,0,0}
 \definecolor{YELLOW}{cmyk}{0,0,1,0}
 }

\@ifundefined{definecolor}{\@ifundefined{definecolor}
 {\@ifundefined{definecolor}
 {\@ifundefined{definecolor}
 {\@ifundefined{definecolor}
 {\@ifundefined{definecolor}
 {\usepackage{color}}{}
}{}
}{}
}{}
}{}
}{}
\makeatother

\begin{document}

\title{Approximate particle number distribution from direct stochastic sampling of the Wigner function}

\author{R.~J.~Lewis-Swan}

\affiliation{The University of Queensland, School of Mathematics and Physics,
Brisbane, Queensland 4072, Australia}

\author{M.~K.~Olsen}

\affiliation{The University of Queensland, School of Mathematics and Physics,
Brisbane, Queensland 4072, Australia}

\author{K.~V.~Kheruntsyan}

\affiliation{The University of Queensland, School of Mathematics and Physics,
Brisbane, Queensland 4072, Australia}

\date{\today }
\begin{abstract}
We consider the Wigner quasi-probability distribution function of a single mode of an electromagnetic or matter-wave field to address the question of whether a direct stochastic sampling and 
binning of the absolute square of the complex field amplitude can yield a distribution function $\tilde{P}_n$ that closely approximates the true particle number probability distribution $P_n$ of the 
underlying quantum state. By providing an operational definition of the binned distribution $\tilde{P}_n$ in terms of the Wigner function, we explicitly calculate the overlap 
between $\tilde{P}_n$ and ${P}_n$ and hence quantify the statistical distance between the two distributions. We find that there is indeed a close quantitative correspondence between $\tilde{P}_n$ 
and $P_n$ for a wide range of quantum states that have smooth and broad Wigner function relative to the scale of oscillations of the Wigner function for the relevant Fock state. 
However, we also find counterexamples, including states with high mode occupation, for which $\tilde{P}_n$ does not closely approximate $P_n$.
\end{abstract}


\maketitle

\section{Introduction}

The Wigner function, or the Wigner quasi-probability distribution \cite{Wigner1932,moyal1949quantum,Leonhardt-QuantumOptics,schleich2011quantum,WallsMilburn}, 
has proven to be a versatile tool in understanding quantum mechanics. Firstly, by providing a complete representation of the quantum mechanical density operator in 
phase space, the Wigner function serves as the quantum moment-generating functional that allows the calculation of quantum mechanical expectation values of operators 
in the spirit of classical statistical physics. Secondly, the Wigner function has been extensively used in the so-called truncated Wigner approximation as a calculation 
technique for quantum dynamical simulations, most notably in the fields of quantum optics and ultracold atoms 
\cite{drummond1993simulation,Werner1995simulation,steel1998dynamical,sinatra00wigner,*sinatraPRL2001,*sinatra02wigner,gardiner2002SGPE,Polkovnikov_Wigner_2003,norrie05turbulence,*norrie06turbulence,Ruostekoski_Isella_PRL_2005,*Isella_Ruostekoski_2005,*Isella_Ruostekoski_2006,deuar2007posp,polkovnikov2010phasespace}. 
This latter utility follows from the possibility of converting the master equation for the quantum density operator into a generalised Fokker-Planck equation, 
which itself -- for dissipationless systems and after truncation of third- and higher-order derivative terms (if any) \cite{JoelMurray:2015} -- 
acquires the form of a classical Liouville equation and can be cast as an equivalent set of stochastic $c$-number differential 
equations for the phase-space variables.

Despite the formal analogy of the evolution equation for the Wigner function to the Liouville equation for a classical probability distribution, the strict 
interpretation of the Wigner function as a true probability distribution fails as it can attain negative values for certain quantum states. Furthermore, even when 
the Wigner function is strictly non-negative, its difference from a classical probability distribution stems from the fact that it is still constrained by the quantum 
mechanical uncertainty principle: it is a joint quasi-probability distribution for quantum mechanically \textit{incompatible} observables and, therefore, cannot be 
regarded as a true probability distribution. In the truncated Wigner approximation (TWA), 
this constraint manifests itself through the fact that even though the $c$-number differential equations formally coincide with their classical 
deterministic counterparts, the quantum mechanical uncertainties are mimicked via random initial conditions that are sampled stochastically from the Wigner-function 
representation of the initial density matrix. Accordingly, the individual stochastic realisations or phase-space trajectories of the complex field amplitude
do not have any correspondence to physical observables, except in the mean where they correspond to expectation values of symmetrically ordered creation and annihilation operators. 


Given this understanding of the auxiliary role of the individual stochastic trajectories, we nevertheless consider a simple practical procedure of directly binning the individual stochastic 
realisations of the absolute square of the complex field amplitude and address the following questions: (i) can the resulting, essentially \textit{heuristic}, distribution be nevertheless operationally 
defined in terms of the Wigner function of the underlying quantum state; and (ii) under what conditions, if any, will this distribution closely approximate the true particle number probability 
distribution function $P_n$?

More specifically, focusing for concreteness on problems involving a non-negative initial Wigner function $W_{| \psi \rangle}(\alpha)$ of state $| \psi \rangle$---such that its non-negativity throughout 
the ensuing dynamics is either intrinsically preserved (such as for systems described by Hamiltonians that depend no-higher-than quadratically on creation and annihilation field operators) or enforced by the 
truncated Wigner approximation \cite{Hudson1974,JoelMurray:2015}---we construct the binned number distribution $\tilde{P}_n$ by calculating $n_i = |\alpha_i|^2 - 1/2$. Here, $\alpha_i$ is the 
complex amplitude of a single-mode field and the index $i$ indicates an individual trajectory (or equivalently individual samples appropriately taken from a known Wigner function). We subsequently
sort the continuous values into discrete bins such that $\tilde{P}_n$ is the probability to find $n_i$ in the interval $n-1/2 \leq n_i < n+1/2$. 
The subtraction of $1/2$ in the calculation of samples of $n_i$ corresponds to the subtraction on average of half a quantum occupation,
which is required in the calculation of the mean mode population 
$\langle \hat{n}  \rangle = \langle \hat{a}^{\dagger} \hat{a} \rangle \equiv \langle \alpha^*\alpha\rangle_W-1/2$ (where $\hat{n} $ is the particle number operator, 
while $\hat{a}^{\dagger} $ and $\hat{a}$ are the mode creation and annihilation operators) to ensure the correspondence of $\langle \alpha^*\alpha\rangle_W$ to a symmetrically-ordered Wigner moment .

Apart from being purely of academic interest, the main questions that we address here have practical implications: even though the true $P_n$ can, in principle, be calculated from the  
$W_{| \psi \rangle}(\alpha)$ (assumed to be known either analytically or reconstructed numerically), the calculation can become computationally very demanding and impractical for highly occupied states (see Sec.~\ref{sec:FormalDeriv}). 
In contrast, constructing the binned distribution $\tilde{P}_n$ is a simple and straightforward procedure. In addition, direct binning of individual stochastic realisations becomes intuitively 
justified in the classical limit, such as in the realm of the classical field method based on the TWA 
\cite{steel1998dynamical,sinatraPRL2001,DavisBlakie2008,polkovnikov2010phasespace,janne10darksoliton,Deuar11solitons_qbec,deuar12solitons_thermal,janne13classicality,janne14measurement}. 
For example, Blakie \textit{et. al.} make the remark that \textit{``For highly occupied fields, the behaviour observed in each trajectory of the TWA seems to be typical of that seen in single realizations 
of experiments. Thus, it is plausible that single realizations of Wigner trajectories should approximately correspond to a possible outcome of a given experiment''}. In this sense, in the realm of the classical 
field method, the procedure of binning the individual stochastic realisations of $|\alpha_i|^2 - 1/2$ becomes similar to acquiring the particle number distribution from the histograms of individual 
experimental runs aimed at particle detection. By offering an operational definition of $\tilde{P}_{n}$, which enables us to quantify its similarity to the true $P_n$, we essentially provide 
a way to quantitively assess such an interpretation of the individual Wigner trajectories.

We find that the defining feature governing the similarity of $\tilde{P}_n$ to the true $P_n$ is the smoothness and the broadness of the Wigner function relative to the oscillatory structure 
in $W_{|n\rangle}(\alpha)$. For some states, e.g, thermal and coherent, this criterion is in fact equivalent to high mode occupation assumed in the classical field method. 
However, we  also show---using an explicit counterexample for a highly squeezed coherent state (the Wigner function of which is always positive and smooth)---that high 
mode occupation alone is not always sufficient for such a similarity and cannot be generally used to assert the `classical'-like nature of the mode in question. 
In contrast, the broadness of the Wigner distribution 
\textit{can} serve as the sufficient condition.

The article is organized such that in Sec. \ref{sec:FormalDeriv} we demonstrate formally the underlying mathematical relation between $P_n$ and $\tilde{P}_n$ in the Wigner representation and the 
conditions on $W_{|\psi\rangle}(\alpha)$ for $\tilde{P}_n$ to approximately correspond to $P_n$. In Sec. \ref{sec:RelationshipP} we investigate quantitatively the legitimacy of the 
method by applying it to the thermal and squeezed coherent states.
Conversely, in Sec. \ref{sec:Breakdown} we examine under what conditions we expect the method to fail, and how such a failure would manifest in calculations by considering 
highly-squeezed states. Finally, in Sec. \ref{sec:BoseHubbard} 
we demonstrate a practical application of our method to a numerical example with an a priori unknown Wigner function, following the criteria of validity outlined and investigated in the prior sections.

\section{Formal Derivation \label{sec:FormalDeriv}}

To formally evaluate the particle number distribution $P_n$ of a single-mode state $|\psi \rangle$, one may calculate the overlap of the state $|\psi\rangle$ with the Fock state $|n\rangle$, 
which in the Wigner representation is given by \cite{Leonhardt-QuantumOptics}
\begin{eqnarray}
 P_n \equiv |\langle \psi | n \rangle|^2 = \pi \int d^2\alpha W_{|\psi\rangle}(\alpha) W_{|n\rangle}(\alpha), \label{eqn:Pn_defn}
\end{eqnarray}
where $W_{|\psi\rangle}(\alpha)$ and $W_{|n\rangle}(\alpha)$ are the respective Wigner functions, with $W_{|n\rangle}(\alpha)$ given by \cite{Leonhardt-QuantumOptics}
\begin{equation}
 W_{|n\rangle}(\alpha) = \frac{2}{\pi}(-1)^n e^{-2|\alpha|^2} L_{n}(4|\alpha|^2), \label{eqn:W_n_exact}
\end{equation}
where $L_n(x)$ is the $n$th-order Laguerre polynomial. With knowledge of the explicit form of $W_{|\psi\rangle}(\alpha)$ one may then analytically or numerically evaluate the 
integral in Eq.~(\ref{eqn:Pn_defn}) to derive the number distribution of the state exactly. In dynamical simulations one may numerically solve the integral (\ref{eqn:Pn_defn}) by first reconstructing the 
Wigner function $W_{|\psi\rangle}(\alpha)$ itself, or by noting that the rhs of Eq.~(\ref{eqn:Pn_defn}) is formally equivalent to
\begin{equation}
 P_n \equiv \pi \langle W_{|n\rangle}(\alpha) \rangle_W , \label{eqn:Pn_WigTrick}
\end{equation}
where the subscript refers to averaging over many stochastic trajectories which provide samples of $\alpha_i$ according to the distribution $W_{|\psi\rangle}(\alpha)$. Such a computation 
is in general non-trivial for highly occupied states or those with a sufficiently broad number distribution as it requires evaluation of high-order Laguerre polynomials with large arguments. 
Usually, such computations require numerical techniques such as quadruple precision to overcome stability issues for $n \gtrsim 330$~\footnote{This is based on a computation of $\mathrm{exp}(-2|\alpha|^2)L_n(4|\alpha|^2)$ 
which fails at $|\alpha|^2\approx360$ and $n=330$ using MATLAB R2013a (double precision) and an algorithm based on the standard recursive definition of $L_n(x)$. Increasing $n$ leads to failure at decreasing values of $|\alpha|^2$.}. 


In contrast, direct binning of individual stochastic trajectories overcomes such computational issues and offers a much simpler method to implement numerically.
To characterize the connection of $\tilde{P}_n$ to the formal definition of $P_n$ we can mathematically define the binned probability distribution as
\begin{equation}
 \tilde{P}_n \equiv \int^{n+1}_{n} d(|\alpha|^2) ~ \mathcal{P}(|\alpha|^2) , \label{eqn:binning_eqn}
\end{equation}
where $\mathcal{P}(|\alpha|^2)$ is the probability density of sampling $|\alpha|^2$ from an ensemble of stochastic trajectories. In terms of the Wigner function, 
this is equivalent to the probability of sampling $\alpha$ from within an annulus in phase-space with inner and outer radii of $\sqrt{n}$ and $\sqrt{n+1}$ respectively. 
Thus we may rewrite Eq.~(\ref{eqn:binning_eqn}), using the Heaviside step function $\theta(x)$, as
\begin{equation}
 \tilde{P}_n  =  \pi \! \int \! d^{2}\alpha \left[\frac{1}{\pi}\theta(|\alpha| - \sqrt{n})\theta(\sqrt{n+1} - |\alpha|) \right] W_{|\psi\rangle}(\alpha). \label{eqn:BohrSommerfeld}
\end{equation}

Comparing now the result of Eq.~(\ref{eqn:BohrSommerfeld}) to Eq.~(\ref{eqn:Pn_defn}) we see that the binning procedure is mathematically equivalent to approximating
$W_{|n\rangle}(\alpha)$ by a radially symmetric boxcar function in phase-space defined as 
\begin{equation}
 \tilde{W}_{|n\rangle}(\alpha) = \frac{1}{\pi}\theta(|\alpha| - \sqrt{n})\theta(\sqrt{n+1} - |\alpha|). \label{eqn:W_n_boxcar}
\end{equation}
This representation of the Fock state Wigner function is known as a Planck-Bohr-Sommerfeld band \cite{schleich2011quantum}, and is equivalent to a smearing out of the classical 
(Kramers) trajectory of a Fock state in phase-space, which is a ring along $|\alpha| = \sqrt{n+1/2}$. The binning procedure as characterized by Eq.~(\ref{eqn:BohrSommerfeld}) is 
then similar to the area-of-overlap 
formalism developed previously by Schleich \cite{schleich2011quantum}, wherein the number distribution of a state can be approximated by the overlap of the phase-space distribution with 
a band in phase-space, representing the number state. We point out the subtle difference that Schleich's formalism can account for interference between probability amplitudes, which is 
equivalent to retaining negative contributions in Eq.~(\ref{eqn:Pn_defn}), whereas the binning procedure rules this out as Eq.~(\ref{eqn:BohrSommerfeld}) is a sum of contributions from a 
strictly non-negative Wigner function. 

\begin{figure}
\includegraphics[width=8.6cm]{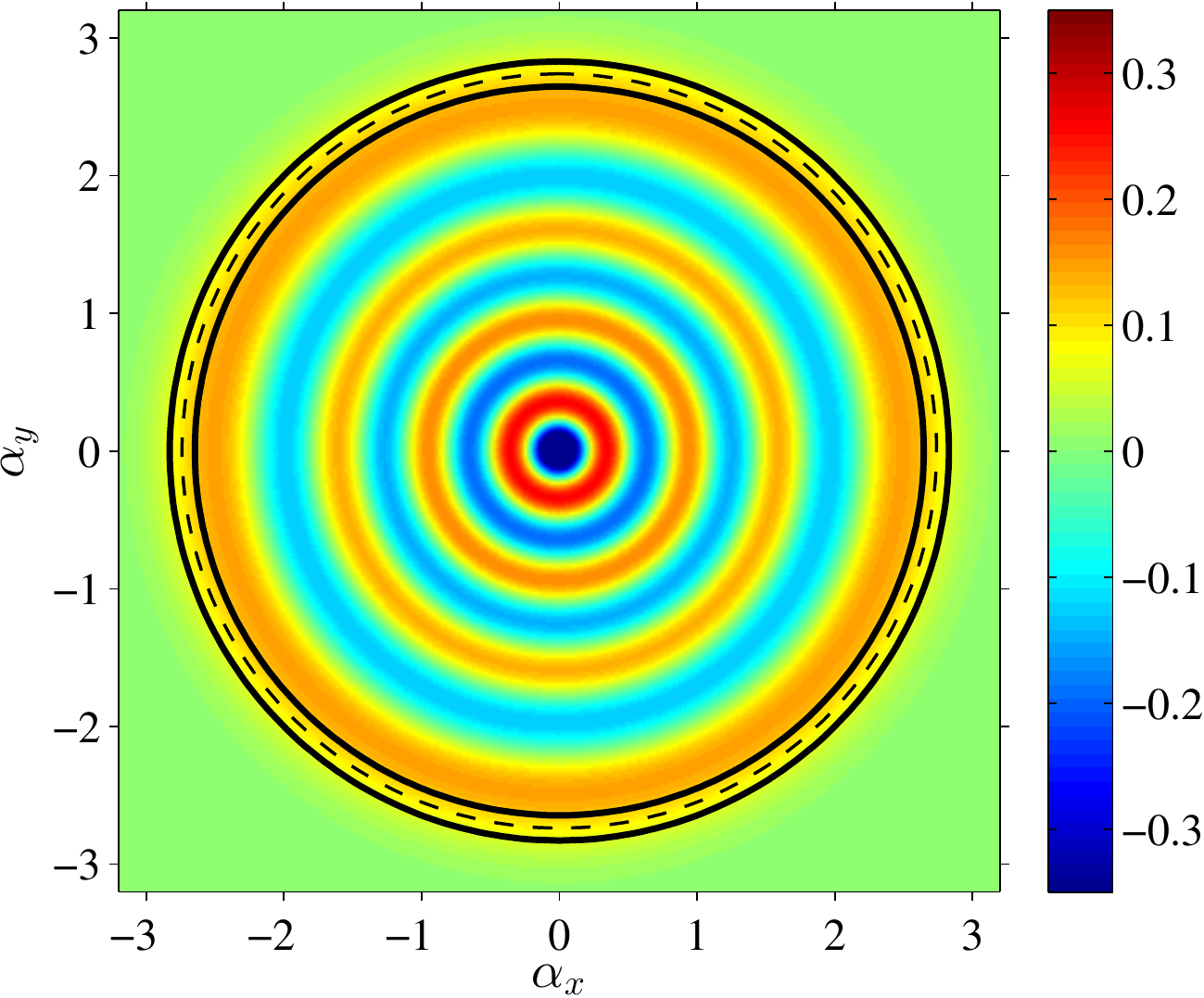}
\caption{Colormap plot of the Wigner distribution, $W_{|n\rangle}(\alpha)$ of the $n=7$ Fock state, Eq.~(\ref{eqn:W_n_exact}), where the axis correspond to $\alpha_x\equiv \mathrm{Re}(\alpha)$ and $\alpha_y\equiv \mathrm{Im}(\alpha)$. The radial oscillations appear distinctly as 
a series of alternating peaks ($W_{|n\rangle}(\alpha)>0$) and troughs ($W_{|n\rangle}(\alpha)<0$). For illustration, we overlay the Planck-Bohr-Sommerfeld band for 
the equivalent state, Eq.~(\ref{eqn:W_n_boxcar}). The inner and outer radii (solid lines) are $\sqrt{n}$ and $\sqrt{n+1}$, which are centered around the 
`classical' trajectory (dashed line) which is a ring of radius $\sqrt{n+1/2}$.}
\label{fig:Fock_pcolor} 
\end{figure}

One can also motivate the approximation of $\tilde{W}_{|n\rangle}(\alpha)$ with a more practical argument, by arguing that low-order moments of $\alpha$ with 
respect to $W_{|n\rangle}(\alpha)$ are dominated by contributions of the final `crest' in the highly-oscillatory Wigner distribution, whilst earlier contributions 
effectively cancel out.
This approach is based on approximations applied by Gardiner \textit{et. al.} in Ref.~\cite{gardiner2002SGPE}, wherein the authors observed that the Wigner 
function of the Fock state could be approximated as a radially symmetric Gaussian ring, 
$\mathcal{W}_{|n\rangle}(\alpha) = \mathcal{A} \;  \exp[ - 2\left( |\alpha|^2 - n - 1/2 \right)^2 ]$ (with $ \mathcal{A} $ being the normalization constant),
which is strictly positive. In Refs.~\cite{olsen2009WignerRep,olsen2004photoassociation} Olsen \textit{et al.} demonstrated explicitly that sampling of 
$\mathcal{W}_{|n\rangle}(\alpha)$ indeed produced all moments $\langle |\alpha|^m \rangle_W$ of the exact Wigner distribution up to $\mathcal{O}(1/n^2)$ 
relative to the leading order, implying that the contribution of all but the final oscillation in $W_{|n\rangle}(\alpha)$ can be considered approximately negligible. 
In light of this, one could also regard $\tilde{W}_{|n\rangle}(\alpha)$, Eq.~(\ref{eqn:W_n_boxcar}), as a further crude approximation to $\mathcal{W}_{|n\rangle}(\alpha)$.

Following the reasoning of Gardiner \textit{et. al.} \cite{gardiner2002SGPE}, we thus intuitively expect the replacement of $W_{|n\rangle}(\alpha)$ by $\tilde{W}_{|n\rangle}(\alpha)$ in 
Eq.~(\ref{eqn:BohrSommerfeld}) to only be a good approximation when $W_{|\psi\rangle}(\alpha)$ is a sufficiently smooth function of $\alpha$ in the region of overlap of the distributions $W_{|\psi\rangle}(\alpha)$ and 
$W_{|n\rangle}(\alpha)$. Qualitatively, due to the radial symmetry of $W_{|n\rangle}(\alpha)$, this means that we require the radial component of $W_{|\psi\rangle}(\alpha)$ to be slowly varying on the order of the characteristic length 
scale $l_{\mathrm{osc}}$ of oscillations in $W_{|n\rangle}(\alpha)$, which, due to the properties of the Laguerre polynomial $L_n(x)$, can be estimated to be $l_{\mathrm{osc}} \sim 1/\sqrt{n}$.

A quantitative form of this criteria can be defined by introducing the radial distribution $w_{|\psi\rangle}(r) = \int d\phi W_{|\psi\rangle}(r,\phi)$, which corresponds to integrating out the angular component of the 
Wigner function, which we have rewritten in terms of polar co-ordinates $r \equiv |\alpha|$ and $\phi \equiv \mathrm{Arg}(\alpha)$. The smoothness of this radial distribution can be quantified by the characteristic 
inhomogeneity length \cite{Kheruntsyan2005},
\begin{equation}
 l^{|\psi\rangle}_{\mathrm{inh}}(r) \equiv \frac{w_{|\psi\rangle}(r)}{\left| \partial w_{|\psi\rangle} / \partial r \right|} .
\end{equation}
In terms of this smoothness measure, the substitution of $\tilde{W}_{|n\rangle}(\alpha)$ in place of $W_{|n\rangle}(\alpha)$ in Eq.~(\ref{eqn:BohrSommerfeld}) for a \emph{specific} $n$ 
requires that
\begin{equation}
 \frac{l^{|\psi\rangle}_{\mathrm{inh}}(r)}{l_{\mathrm{osc}}} \simeq \sqrt{n} l^{|\psi\rangle}_{\mathrm{inh}}(r) \gg 1 , \label{eqn:SmoothnessCriteria}
\end{equation}
for all $r$ in the region of overlap of the radially symmetric $W_{|n\rangle}(r) \equiv W_{|n\rangle}(\alpha)$ and $w_{|\psi\rangle}(r)$, i.e. those which will contribute to the integral of Eq.(\ref{eqn:Pn_defn}). 
This region can be determined in a straightforward manner by noting that $W_{|n\rangle}(r)$ will be contained within the region $0 \leq r \lesssim \sqrt{n+1}$ (see Fig.~\ref{fig:Fock_pcolor} for an example of this for $n=7$), 
whilst $w_{|\psi\rangle}(r)$ can be constructed either analytically from a known $W_{|\psi\rangle}(\alpha)$ or numerically for easy comparison. Consequentially, a sufficient condition for the 
complete distribution $\tilde{P}_n$ to be a valid approximation to $P_n$ for \emph{all} $n$ is that Eq.~(\ref{eqn:SmoothnessCriteria}) is satisfied for the smallest relevant $n$ (hence largest length scale $l_{\mathrm{osc}} \sim 1/\sqrt{n}$) 
for all $r$ where the distribution $w_{|\psi\rangle}(r)$ is appreciable.

There are two complementary properties of $W_{|\psi\rangle}(\alpha)$ which we qualitatively expect to satisfy the smoothness criterion of Eq.~(\ref{eqn:SmoothnessCriteria}). 
Firstly, for states localized near the origin in phase-space -- such as the thermal state -- one requires that the Wigner function has a characteristic radial width 
$\sigma \gg 1$. This implies that $l^{|\psi\rangle}_{\mathrm{inh}}(r) \gg 1$ and thus $\tilde{P}_n$ will approximate $P_n$ well even for small $n \sim 1$. Secondly, for states of fixed width -- such as the coherent state -- 
one requires a large coherent displacement $|\beta|$ from the origin. As the overlap between $W_{|\psi\rangle}(\alpha)$ and $W_{|n\rangle}(\alpha)$ will 
generally be greatest for $n\sim |\beta|^2$, the length-scale of the oscillations in $W_{|n\rangle}(\alpha)$ in the relevant regions of $W_{|\psi\rangle}(\alpha)$ will scale 
as $1/|\beta|$. The radial width of $W_{|\psi\rangle}(\alpha)$ -- and thus the characteristic length scale $l^{|\psi\rangle}_{\mathrm{inh}}$ -- relative to the scale of these oscillations 
thus increases as $|\beta|$ increases, improving the validity of replacing $W_{|n\rangle}(\alpha)$ with $\tilde{W}_{|n\rangle}(\alpha)$. In the following section we illustrate these arguments both 
qualitatively and quantitatively for the thermal and broader class of squeezed coherent states.


Lastly, although this derivation has focused on the single-mode case it may be trivially generalized to a multi-mode state and an equivalent form of $\tilde{P}_{n_1,n_2...}$ may be found. 
The same generalized conditions regarding the relative width of the Wigner function may be applied. However, in the following section we will continue to focus our 
analysis on the single-mode case as it allows us to illustrate the correspondence between the two 
distributions in a transparent manner.

\section{Similarity of $P_n$ and $\tilde{P}_n$ \label{sec:RelationshipP}}

\subsection{Thermal state}
The first state we consider is the thermal state, which is a mixed state defined by the density matrix 
\begin{equation}
 \hat{\rho}_{\mathrm{th}} = \sum^{\infty}_{n=0} P_n | n \rangle \langle n | ,
\end{equation}
where the number distribution is given by \cite{WallsMilburn}
\begin{equation}
 P_n = \frac{\bar{n}^n}{(\bar{n} + 1)^{n+1}} ,
\end{equation}
and is characterized solely by the mean occupation $\langle \hat{n} \rangle = \bar{n}$. 

The corresponding Wigner function is \cite{WallsMilburn}
\begin{equation}
 W_{\mathrm{th}}(\alpha) = \frac{1}{\pi(\bar{n}+1/2)} \mathrm{exp}\left(- \frac{|\alpha|^2}{\bar{n}+1/2} \right) ,
\end{equation}
which has a radial rms width $\sigma = \sqrt{(\bar{n} + 1/2)/2}$. Therefore, according to our criterion, the sufficient requirement ($\sigma \!\gg \!1$) for $\tilde{P}_n$ to agree well 
with the physical $P_n$ is equivalent in this case to high mean mode occupation $\bar{n} \!\gg \!1$.

Substituting $W_{\mathrm{th}}(\alpha)$ into Eq.~(\ref{eqn:BohrSommerfeld}) leads to
\begin{equation}
 \tilde{P}_n = e^{-n/(\bar{n}+1/2) } \left[ 1 - e^{ -n/(\bar{n}+1/2) } \right] .
\end{equation}
Although this form of $\tilde{P}_n$ clearly differs from $P_n$, a keen eye will note that in fact 
\begin{equation}
 \tilde{P}_n = \frac{\langle n \rangle_{\mathrm{bin}}^n}{(\langle n \rangle_{\mathrm{bin}} + 1)^{n+1}} ,
\end{equation}
where 
\begin{equation}
 \langle n \rangle_{\mathrm{bin}} \equiv \sum_{n=0}^{\infty} n\tilde{P}_n = \frac{1}{e^{1/(\bar{n}+1/2)} - 1} .
\end{equation}
Hence while both distributions may be written solely in terms of their respective means, $\tilde{P}_n \neq P_n$ explicitly as $\langle n \rangle_{\mathrm{bin}} \neq \bar{n}$. 

To estimate the applicability of $\tilde{P}_n$ we analytically evaluate the characteristic inhomogeneity length of the radial distribution 
$w_{\mathrm{th}}(r)$ to be
\begin{equation}
 l^{\mathrm{th}}_{\mathrm{inh}}(r) = \frac{\bar{n} + 1/2}{2r} .
\end{equation}
As previously discussed, for a specific Fock state $|n\rangle$ the Wigner function $W_{|n\rangle}(r)$ is only appreciable in the region $0 \leq r \lesssim \sqrt{n}$, which implies that 
the region of overlap of $w_{\mathrm{th}}$ and $W_{|n\rangle}(r)$ has an upper bound of $r \lesssim \sqrt{n+1}$. Using this restriction, we find a lower bound on the characteristic 
inhomogeneity length within this region of $l^{\mathrm{th}}_{\mathrm{inh}} \lesssim (\bar{n} + 1/2)/2\sqrt{n+1}$. The condition $\sqrt{n}l^{\mathrm{th}}_{\mathrm{inh}} \gg 1$ [Eq.~(\ref{eqn:SmoothnessCriteria})] then reduces to the 
requirement $\bar{n} \gg 1$ for arbitrary $n$. That $\tilde{P}_n$ is valid for $\bar{n} \gg 1$ is an intuitive result as this corresponds to a broad radial rms width 
$\sigma \approx \sqrt{\bar{n}/2} \gg 1$ of the Wigner function $W_{\mathrm{th}}(\alpha)$.


As a quantitative measure of how well the binned particle number distribution $\tilde{P}_n$ approximates the true distribution $P_n$,
we use the Bhattacharyya statistical distance
\cite{DB_paper}
\begin{equation}
 D_B = -\mathrm{ln}[B(P, \tilde{P})] , \label{eqn:DB_defn}
\end{equation}
where the Bhattacharyya coefficient is given by
\begin{equation}
 B(P, \tilde{P}) = \sum^{\infty}_{n = 0} \sqrt{P_n \tilde{P}_n}.   
\end{equation}
For $\tilde{P}_n \!\rightarrow \!P_n$ the Bhattacharyya coefficient becomes $B(P,\tilde{P}) \!\rightarrow \!\sum^{\infty}_{n = 0} P_n  \!= \!1$ due to the normalization condition and 
hence $D_B \rightarrow 0$, indicating complete overlap of the distributions. 

For the thermal state the Bhattacharyya coefficient can be calculated exactly to give
\begin{equation}
 B(P, \tilde{P}) = \frac{\left[ 1 - e^{-2/(2\bar{n}+1)} \right]^{1/2}}{\left( \bar{n}+1 \right)^{1/2} - \bar{n}^{1/2}e^{-1/(2\bar{n}+1)}} ,
\end{equation}
and thus the Bhattacharyya distance is
\begin{eqnarray}
 D_B & = & -\frac{1}{2}\mathrm{ln}\left[ 1 - e^{-2/(2\bar{n}+1)}  \right] \notag \\
 & & + \mathrm{ln}\left[ \sqrt{\bar{n}+1} - \sqrt{\bar{n}}e^{-1/(2\bar{n}+1)} \right]. \label{eqn:DB_th}
\end{eqnarray}

\begin{figure}
\includegraphics[width=8.6cm]{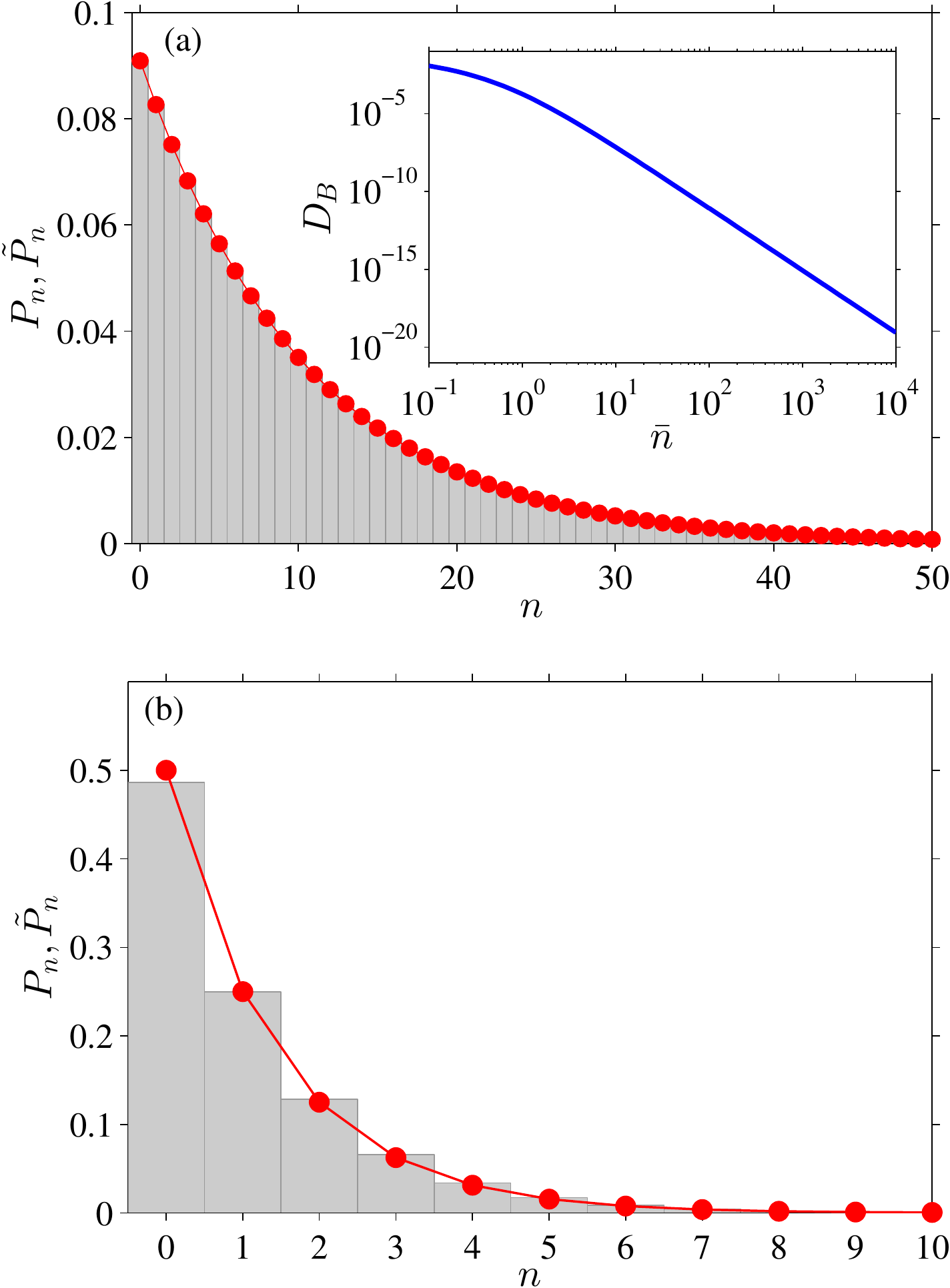}	
\caption{Comparison of the true particle number distribution $P_n$ (red markers) with the binned number distribution $\tilde{P}_n$ (grey bars), for a thermal state of: (a) $\bar{n} = 10$ 
and (b) $\bar{n} = 1$. Data points for $P_n$ are connected for visual clarity. In the inset we plot the statistical distance $D_B$ between the two distributions, Eq.~(\ref{eqn:DB_th}), for a range of mean 
occupations $\bar{n}$, which scales as $\propto 1/\bar{n}^4$ for $\bar{n}\gg 1$.
}
\label{fig:DB_plot_th} 
\end{figure}

In the limit of $\bar{n} \gg 1$ we find the behaviour 
\begin{equation}
 D_B \propto \bar{n}^{-4}, \label{eqn:DB_th_scaling}
\end{equation}
which can be recast in terms of the width of the Wigner function $W_{\mathrm{th}}(\alpha)$, $\sigma \simeq \sqrt{\bar{n}/2}$ for $\bar{n} \gg 1$, as
\begin{equation}
 D_B \propto \sigma^{-8}.
\end{equation}
This strong scaling clearly shows that for large mean occupation $\bar{n} $, or equivalently for a sufficiently broad Wigner function, the binned distribution $\tilde{P}_n$ rapidly approaches 
the true $P_n$. To illustrate this, we plot a comparison of the two distributions for a thermal state of $\bar{n} = 10$ and $\bar{n}=1$ in Fig.~\ref{fig:DB_plot_th}. As we see, even only moderately large 
mean occupations, such as $\bar{n}=10$, render the two distributions nearly identical (quantitatively, the largest absolute discrepancy is $\simeq 6.5\times10^{-5}$ for $n=0$). Moreover, we find good agreement is retained for states with a 
population as small as $\bar{n}=1$ (in this case the worst absolute discrepancy is $\simeq 0.01$ for $n=0$). 



\subsection{Squeezed coherent state \label{sec:SqueezedCoh}}

\begin{figure}
\includegraphics[width=3.9cm]{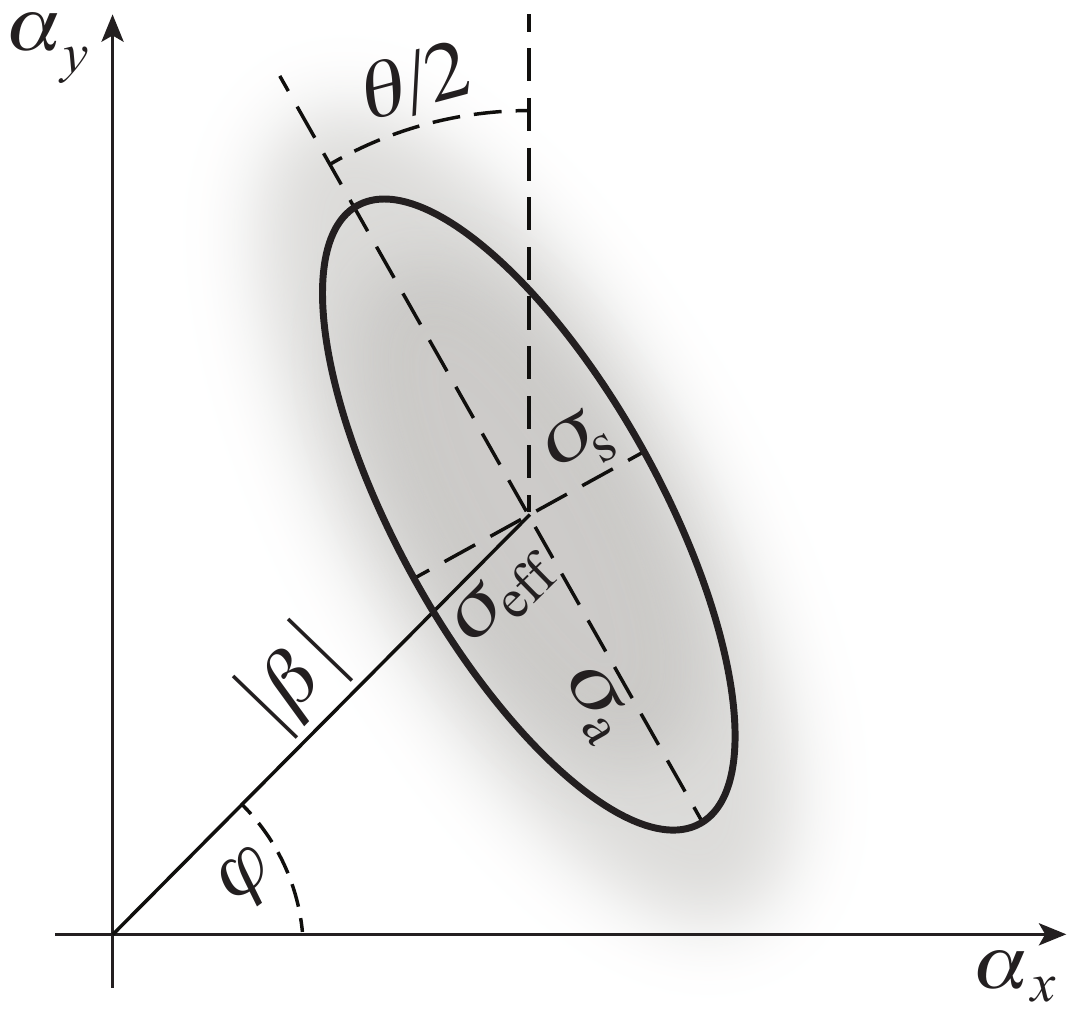}
\caption{Illustration of the Wigner function for a squeezed coherent state $W_{|\beta,\eta\rangle}(\alpha)$. 
The action of the squeezing operator $\hat{S}(\eta)$ on 
the initial state $|0\rangle$ is to squeeze the vacuum state Wigner function (a symmetric Gaussian with rms width $\sigma=1/2$) by $e^{-s}$ along the $\alpha_x$-axis and stretch it by $e^{s}$ along 
the $\alpha_y$-axis, then rotate the distribution by $\theta/2$. The subsequent action of the displacement operator $\hat{D}(\beta)$ is to shift the distribution by $\beta=|\beta|e^{i\varphi}$. 
The relevant length scale in comparison to the radially-directed oscillations in $W_{|n\rangle}(\alpha)$ is the effective width $\sigma_{\mathrm{eff}}$ along the radial direction of $W_{|\beta,\eta\rangle}(\alpha)$.
}
\label{fig:BallStick} 
\end{figure}

The second state which we consider is the squeezed coherent state, defined as
\begin{equation}
 |\beta,\eta\rangle = \hat{D}(\beta) \hat{S}(\eta)|0\rangle ,
\end{equation}
where $\hat{D}(\beta) = \mathrm{exp}(\beta\hat{a}^{\dagger} - \beta^*\hat{a})$ is the displacement operator and the squeezing operator is 
$\hat{S} = \mathrm{exp}[\{\eta^*\hat{a}^2 - \eta(\hat{a}^{\dagger})^2\}/2]$ where $\eta = se^{i\theta}$ for $s\geq 0$ \cite{KnightLoudon87,WallsMilburn}.
In Fig.~\ref{fig:BallStick} we illustrate the actions of these operators in phase-space. Firstly the squeezing operator `squeezes' the Gaussian Wigner 
distribution of the vacuum by an amount $e^{-s}$ along an axis defined by the squeezing angle $\theta$, whilst the perpendicular axis is stretched by $e^{s}$. 
The displacement operator then shifts the distribution in phase space by $\beta = |\beta|e^{i\varphi}$. There exist two special sub-cases of the squeezed coherent state: 
(i) the coherent state $|\beta\rangle$ where $\beta \neq 0$ and $s = 0$; and (ii) the squeezed vacuum state $|0,\eta\rangle$ where $\beta = 0$ and $s \neq 0$.

The Wigner function of the general squeezed coherent state can be written in a simple form \cite{Kim-Knight-1989}
\begin{equation}
 W_{|\beta,\eta\rangle}(\gamma) = \frac{2}{\pi} \mathrm{exp}\left(-\frac{\gamma^2_x}{2\sigma^2_s} - \frac{\gamma^2_y}{2\sigma^2_a}\right) , \label{eqn:Wig_sq_coh}
\end{equation}
where 
 \begin{eqnarray}
 \gamma_x & = & \left(\alpha_x - \beta_x \right)\mathrm{cos}\left(\frac{\theta}{2}\right) + \left(\alpha_y - \beta_y \right)\mathrm{sin}\left(\frac{\theta}{2}\right) , \\
 \gamma_y & = & -\left(\alpha_x - \beta_x \right)\mathrm{sin}\left(\frac{\theta}{2}\right) + \left(\alpha_y - \beta_y \right)\mathrm{cos}\left(\frac{\theta}{2}\right) ,
\end{eqnarray}
for $\alpha = \alpha_x + i\alpha_y$ and $\beta = \beta_x + i\beta_y$. The rms widths along the squeezed and anti-squeezed axes are given by $\sigma_s = e^{-s}/2$ and 
$\sigma_a = e^s/2$, respectively. Independent control over the parameters $\beta$ and $\eta$ allows us to quantitatively probe the similarity of $\tilde{P}_n$ and $P_n$ 
 as a function of the width of the Wigner distribution.

The number distribution of the squeezed state is nontrivial, 
\begin{eqnarray}
 P_n & = & \frac{\left(\frac{1}{2}\mathrm{tanh}(s)\right)^n}{n!\mathrm{cosh}(s)} e^{-|\beta|^2 \left[ 1 + \mathrm{cos}(2\varphi - \theta)\mathrm{tanh}(s) \right] }\notag \\
  & & \times \left| H_n \left( \frac{\beta + \beta^*e^{i\theta}\mathrm{tanh}(s)}{\sqrt{2e^{i\theta}\mathrm{tanh}(s)}} \right) \right|^2 ,
\end{eqnarray}
with mean occupation $\langle \hat{n} \rangle = |\beta|^2 + \mathrm{sinh}^2(s)$ \cite{Yuen1976,KnightLoudon87}. For large coherent displacement such that $|\beta|^2 \gg e^{2s}$, 
this $P_n$ can be approximated by a simple Gaussian \cite{KnightLoudon87}
\begin{equation}
 P_n \simeq \frac{1}{\sqrt{2\pi\langle\Delta^2\hat{n}\rangle}} \mathrm{exp}\left[ \frac{-(n-|\beta|^2)^2}{2\langle\Delta^2\hat{n}\rangle} \right] , \label{eqn:Pn_sq_coh_gaussian}
\end{equation}
whose rms width is given by $\sigma=\sqrt{  \langle\Delta^2\hat{n}\rangle }$, where
\begin{equation}
 \langle\Delta^2\hat{n}\rangle  = |\beta|^2 \left[ e^{-2s}\mathrm{cos}^2\left(\varphi - \frac{\theta}{2}\right)  + e^{2s}\mathrm{sin}^2\left(\varphi - \frac{\theta}{2}\right) \right] .
\end{equation}
This form demonstrates how the squeezing operator stretches or squeezes the probability distribution $P_n$ according to the relative orientation of the squeezing and coherent displacement. 
In this section, our analysis will be limited to a range of squeezing such that the above approximation for $P_n$ is valid. The effects of stronger squeezing and its implications 
for both $P_n$ and $\tilde{P}_n$ will be discussed in Sec. \ref{sec:Breakdown}.

%

An analytic form of $\tilde{P}_n$ can be found by substituting Eq.~(\ref{eqn:Wig_sq_coh}) into the definition of Eq.~(\ref{eqn:BohrSommerfeld}), however, the result 
is not particularly insightful. We point the interested reader to Ref.~\cite{Gilliland62} as a guide to the general form of the calculation. Instead, we numerically 
evaluate $\tilde{P}_n$ by stochastically sampling $W_{|\beta,\eta\rangle}(\alpha)$ according to the prescription of Ref.~\cite{olsen2009WignerRep} and binning the 
calculated occupation of each sample. Such a construction is equivalent to obtaining the same state and results via a dynamical simulation of stochastic 
equations (trajectories) in the Wigner representation, as the phenomenological squeezed vacuum state can be generated from a Hamiltonian for spontaneous parametric down-conversion 
(in the undepleted pump approximation) $\hat{H} = i\hbar[ g^*\hat{a}^2 - g(\hat{a}^{\dagger})^2 ]$, in which case the squeezing parameter $\eta$ is given by 
$\eta \equiv gt$. The subsequent coherent displacement of the squeezed state is achieved by coupling the mode $\hat{a}$ to a classical field of amplitude $\varepsilon$, equivalent to 
evolution under the Hamiltonian $\hat{H} = i\hbar \kappa [\varepsilon^*\hat{a} - \varepsilon\hat{a}^{\dagger} ]$ where $\kappa$ is the coupling strength and hence the resulting 
displacement is related as $\beta \equiv \kappa\varepsilon t$.


To estimate under what conditions we expect $\tilde{P}_n$ to be similar to $P_n$ we evaluate the characteristic inhomogeneity length of the radial distribution $w_{|\beta,\eta\rangle}(r)$. 
For weakly squeezed coherent states with a large displacement, $|\beta| \gg 1$, the characteristic inhomogeneity length is approximately~\footnote{We extract the characteristic inhomogeneity length 
by approximating the radial distribution as $w_{|\beta,\eta\rangle}(r) = \int d\phi W_{|\beta,\eta\rangle}(r,\phi) \approx (1/\alpha_x)\int_{-\infty}^{\infty} d\alpha_y W_{|\beta,\eta\rangle}(\alpha)$  
for a purely real coherent displacement ($\varphi = 0$) and making the replacement $r \equiv \alpha_x$. This is valid for $\beta_x \gg 1$ and weak squeezing $\beta_x \gg e^{2s}$.
This can be generalized trivially for arbitrary coherent displacement.}
\begin{equation}
 l^{|\beta,\eta\rangle}_{\mathrm{inh}}(r) \simeq  \frac{\sigma^2_{\mathrm{eff}}}{\left| r - |\beta| \right|} , \label{eqn:l_sqcoh}
\end{equation}
where
\begin{equation}
 \sigma_{\mathrm{eff}} = \sqrt{\sigma^2_s\mathrm{cos}^2\left( \varphi - \frac{\theta}{2} \right) + \sigma^2_a\mathrm{sin}^2\left( \varphi - \frac{\theta}{2} \right)} , \label{eqn:sigma_eff}
\end{equation}
is the characteristic rms width of $w_{|\beta,\eta\rangle}(r)$, or equivalently the effective radial width of the distribution $W_{|\beta,\eta\rangle}$ (see Fig.~\ref{fig:BallStick}). 
As $w_{|\beta,\eta\rangle}(r)$ will be strongly peaked around $r \sim |\beta| \pm \sigma_{\mathrm{eff}}$ we can bound 
Eq.~(\ref{eqn:l_sqcoh}) in this region as $l^{|\beta,\eta\rangle}_{\mathrm{inh}}(r) \geq \sigma_{\mathrm{eff}}$, and thus the smoothness criteria of 
Eq.~(\ref{eqn:SmoothnessCriteria}) becomes $\sqrt{n}l^{|\beta,\eta\rangle}_{\mathrm{inh}}(r) = \sqrt{n}\sigma_{\mathrm{eff}} \gg 1$. As the number distribution of a weakly squeezed coherent state 
is strongly peaked around $n\sim|\beta|^2$ [see Eq.(\ref{eqn:Pn_sq_coh_gaussian})] this criteria reduces to $|\beta|\sigma_{\mathrm{eff}} \gg 1$ and thus will be satisfied for states 
with sufficiently large displacement such that $|\beta| \gg 1/\sigma_{\mathrm{eff}}$. 

%


\begin{figure}
\includegraphics[width=8.6cm]{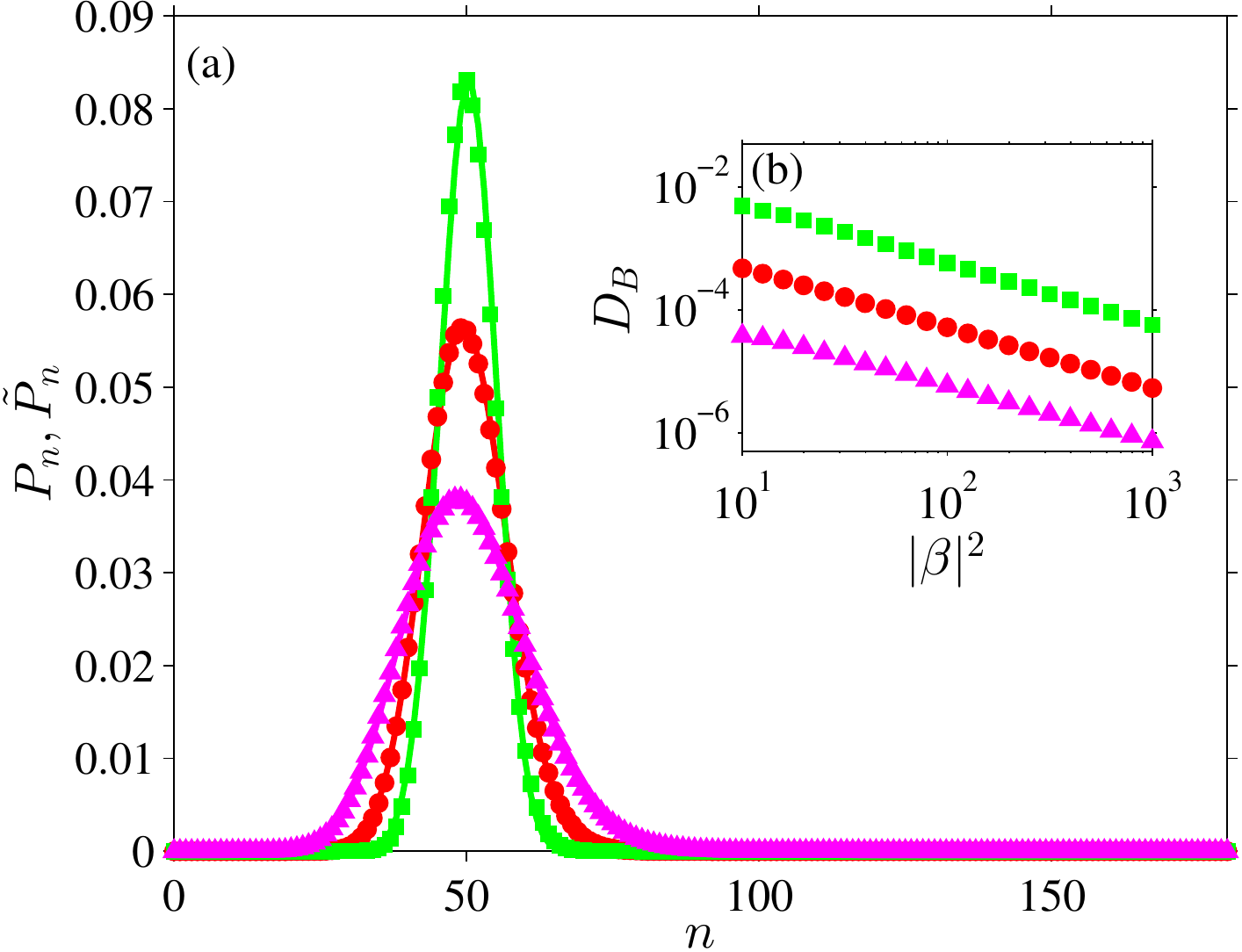}
\caption{
(a) Examples of probability distributions $\tilde{P}_n$ (markers) and $P_n$ (lines) for squeezed coherent ($s\neq 0$) and pure coherent ($s=0$) states, for $|\beta|^2 = 50$. For squeezed coherent states we chose $\varphi = 0$ (see Fig.~\ref{fig:BallStick}), $s = 0.4$, and the squeezing angles of $\theta = 0$ (magenta triangles) and $\theta = \pi$ (green squares); red circles correspond to pure coherent states. (b) Comparison of the respective Bhattacharyya distances as a function of $|\beta|^2$ showing a generic scaling of $D_B \propto |\beta|^{-2}$ for $|\beta|^2 \gg e^{2s}$. Stochastic sampling error of one standard deviation is not indicated but is less 
than $2\%$ of calculated $D_B$ for all data points (obtained from approximately $10^9$ trajectories).
}
\label{fig:DB_sqcoh_plot} 
\end{figure}

In Fig.~\ref{fig:DB_sqcoh_plot}~(a) we plot examples of $\tilde{P}_n$ and $P_n$ for squeezed coherent states with $|\beta|^2 = 50$, $\varphi = 0$, $s = 0.4$, and squeezing angles of 
$\theta = 0$ and $\theta = \pi$, which are referred to as amplitude- and phase-squeezing, respectively. Also plotted is the simple case of a pure coherent state with $s = 0$. 
As we see, the calculated distributions $\tilde{P}_n$ and $P_n$ are visually indistinguishable from each other. The respective Bhattacharyya distances  as a function of 
$|\beta|^2$ are plotted in Fig.~\ref{fig:DB_sqcoh_plot}~(b), where we find a generic scaling independent of $s$, 
\begin{eqnarray}
 D_B \propto |\beta|^{-2},
\end{eqnarray}
in the regime where $|\beta|^2 \gg e^{2s}$ and the approximate form of Eq.~(\ref{eqn:Pn_sq_coh_gaussian}) is valid. This result implies a rapid convergence of 
$\tilde{P}_n$ to $P_n$ with increasing occupation $\langle \hat{n} \rangle \simeq |\beta|^2$.


Beyond the scaling with coherent displacement, we may also quantitatively examine how the absolute width of the Wigner function affects the statistical agreement of 
$\tilde{P}_n$ with $P_n$ by manipulation of the squeezing strength $s$ and angle $\theta$. As highlighted by the discussion of Eq.~(\ref{eqn:l_sqcoh}), we expect the validity of $\tilde{P}_n$ to improve 
as the effective radial width $\sigma_{\mathrm{eff}}$ of the distribution (see Fig.~\ref{fig:BallStick}), $\sigma_{\mathrm{eff}}$
increases with respect to the radially directed oscillations in $W_{|n\rangle}(\alpha)$.

\begin{figure}[tbp]
\includegraphics[width=8.6cm]{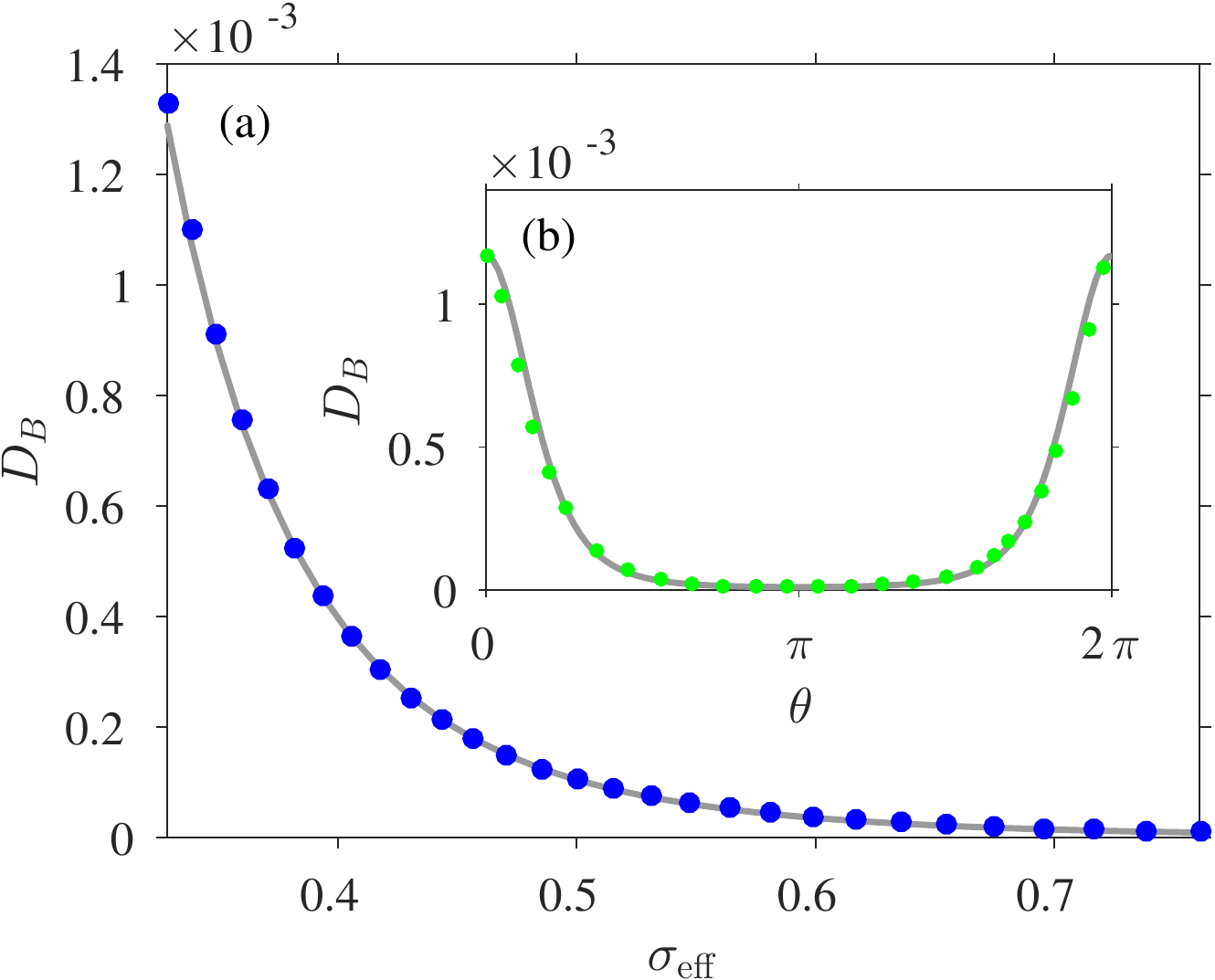}
\caption{(a) Behaviour of statistical distance $D_B$ with the effective width $\sigma_{\mathrm{eff}}$ for a squeezed coherent state with $|\beta|^2 = 50$ and $\varphi = 0$. 
For $\sigma_{\mathrm{eff}} \leq 1/2$, we calculate $D_B$ by fixing the squeezing angle as $\theta = 0$ and thus $\sigma_{\mathrm{eff}} \equiv \sigma_s \leq 1/2$; 
for $\sigma_{\mathrm{eff}} \geq 1/2$, we fix the squeezing angle as $\theta = \pi$ and thus $\sigma_{\mathrm{eff}} \equiv \sigma_a \geq 1/2$. 
A fit $D_B \propto \sigma_{\mathrm{eff}}^{-6}$ (grey line) is also plotted for comparison with the actual stochastically sampled data (blue circles).
(b) Variation of $D_B$ with squeezing angle $\theta$ for a squeezed coherent state with $|\beta|^2 = 50$, $\varphi = 0$ and $s=0.4$ (green circles). 
The behaviour fits the model of Eq.~(\ref{eqn:DB_s_scale}) (grey line) where $\sigma_{\mathrm{eff}}$ depends on the squeezing angle $\theta$ as per Eq.~(\ref{eqn:sigma_eff}).
For numerically calculated data in both (a) and (b) stochastic sampling error of one standard deviation is less than $2\%$ of calculated value (obtained from approximately $10^9$ trajectories).
}
\label{fig:DB_squeezing_phi} 
\end{figure}

We plot the dependence of the Bhattacharyya distance as a function of this parameter in Fig.~\ref{fig:DB_squeezing_phi}~(a) and find it scales as
\begin{equation}
 D_B \propto \sigma^{-6}_{\mathrm{eff}}, \label{eqn:DB_s_scale}
\end{equation}
independently of 
$|\beta|$. This strong scaling again agrees with our intuitive argument, indicating that $\tilde{P}_n$ rapidly 
approaches $P_n$ as the Wigner function becomes increasingly smooth on the length scale of oscillations in $W_{|n\rangle}(\alpha)$. 
The dependence of $D_B $ on the squeezing angle $\theta$ alone is plotted in Fig.~\ref{fig:DB_squeezing_phi}~(b).



\section{Breakdown of relationship \label{sec:Breakdown}}

The analysis of the previous section has demonstrated how, in general, $\tilde{P}_n$ closely replicates $P_n$ when the radial width of the 
Wigner distribution $W_{|\psi\rangle}(\alpha)$ is large compared to the oscillation period of the Fock state Wigner function, $W_{|n\rangle}(\alpha)$. 
If this condition is not satisfied the correspondence breaks down, as we illustrate in this section with two simple counter-examples.
In particular, we demonstrate this with states that are highly-occupied, showing that large occupation alone is not sufficient for approximating $P_n$ by $\tilde{P}_n$.

\begin{figure}[tbp]
\includegraphics[width=8.0cm]{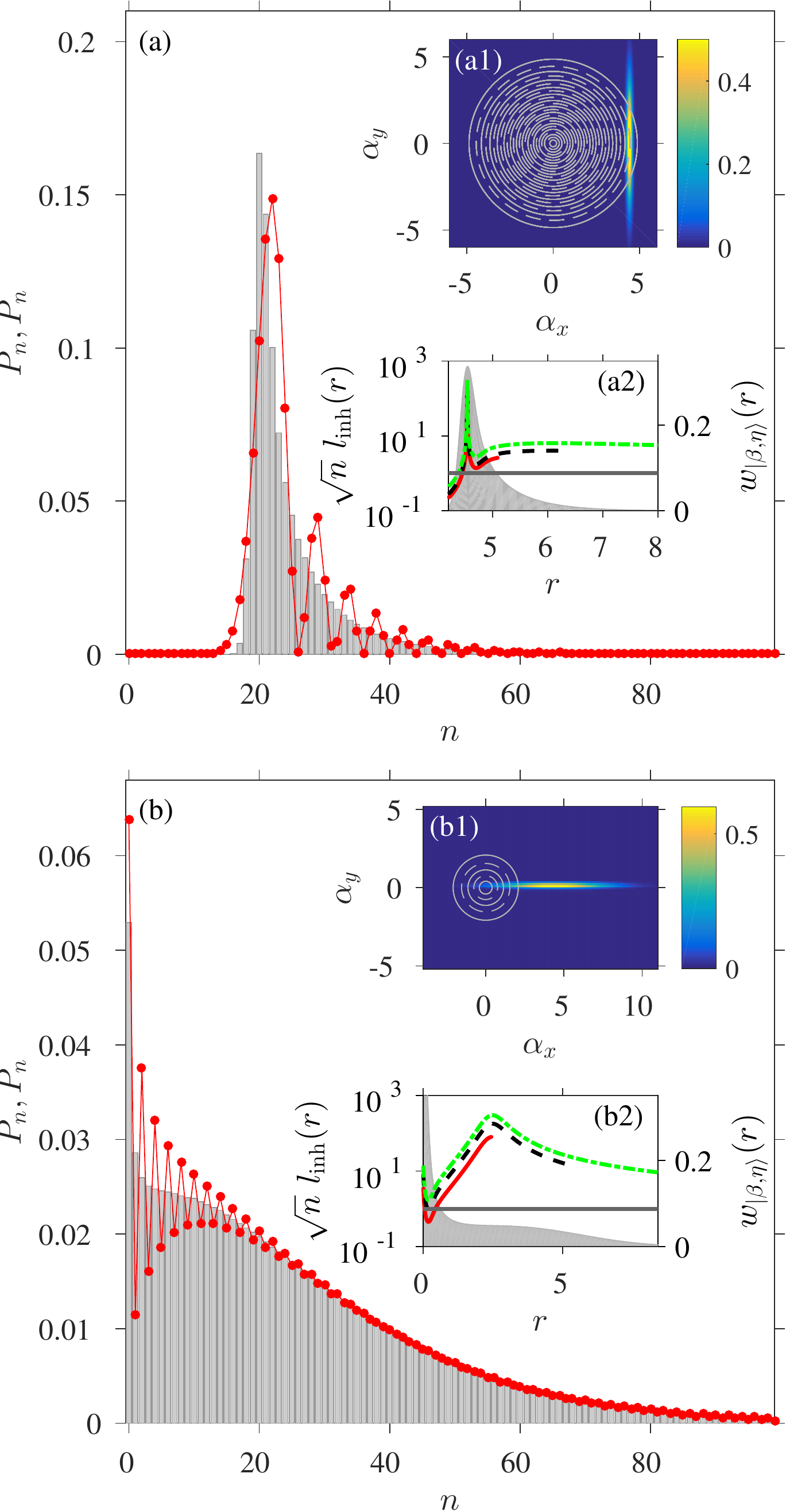}
\caption{Probability distributions $\tilde{P}_n$ and $P_n$  for squeezed coherent states with $|\beta|^2 = 20$ ($\varphi=0$), $s = 1.5$, and squeezing angles of $\theta = 0$ (a) and $\theta = \pi$ (b). 
In (a), the true distribution $P_n$ (red markers) displays oscillations for the full range of $n$, which are not replicated by $\tilde{P}_n$ (grey bars); in (b), the oscillations in $P_n$ occur for $n \lesssim 25$ 
and again are not replicated by $\tilde{P}_n$. Data points for $P_n$ are connected for visual clarity.
The insets (a1) and (b1) show the Wigner functions $W_{|\beta,\eta\rangle}(\alpha)$ of the respective squeezed coherent states. The contour lines represent the Wigner functions $W_{|n\rangle}(\alpha)$ of 
Fock states with $n=25$ and $n=5$, illustrating the maxima (solid lines) and minima (dashed lines) of the corresponding functions. 
In the insets (a2) and (b2) we plot the quantity $\sqrt{n}l^{|\beta,\eta\rangle}_{\mathrm{inh}}(r)$ for: (a2) $n=25$ (red solid line), $n=40$ (dashed black line) and $n=100$ (dot-dashed green line), 
and (b2) $n=5$ (red solid line), $n=25$ (dashed black line) and $n=70$ (dot-dashed green line). Comparison to $\sqrt{n}l^{|\beta,\eta\rangle}_{\mathrm{inh}}(r) = 1$ is indicated (dark grey line). To assist in identifying 
the region of overlap we terminate the lines at $r = \sqrt{n+1}$ to indicate the region containing $W_{|n\rangle}(r)$, whilst underlaying the radial distribution $w_{|\beta,\eta\rangle}(r)$ (shaded grey region) for comparison. 
For both states, we demonstrate that $\sqrt{n}l^{|\beta,\eta\rangle}_{\mathrm{inh}}(r) \gg 1$ in the relevant region of $r$ is only true for: (a2) $n \gg 100$ and (b2) $n \gg 25$, consistent with the poor 
agreement between $\tilde{P}_n$ and $P_n$ for $n \lesssim 25$ and $n \lesssim 100$ in the respective examples. 
}
\label{fig:Pn_squeezebreak} 
\end{figure}

As an example, in Fig.~\ref{fig:Pn_squeezebreak} we plot 
$\tilde{P}_n$ and $P_n$  for $|\beta|^2 \!= \!20$, $s \!= \!1.5$ and for two squeezing angles: (a) $\theta \!= \!0$ and (b) $\theta \!= \!\pi$. In both cases we see a 
range of $n$ emerges where the true probability distribution $P_n$ 
oscillates strongly. 
In terms of the binning procedure, it is clear that $W_{|\beta,\eta\rangle}(\alpha)$ is sufficiently 
elongated---in the region of relevant $r$---that it is approximately the width of the oscillations in $W_{|n\rangle}(\alpha)$ and multiple oscillations become important in the calculation of 
the integral in Eq.~(\ref{eqn:Pn_defn}) as illustrated in Figs.~\ref{fig:Pn_squeezebreak}~(a1) and (b1). 
This is quantitatively supported by examining the characteristic inhomogeneity length $l^{|\beta,\eta\rangle}_{\mathrm{inh}}(r)$ for the respective states, which is plotted in Figs.~\ref{fig:Pn_squeezebreak}~(a2) and (b2).


In both cases, the narrowness and location of the Wigner distribution $W_{|\beta,\eta\rangle}(\alpha)$ implies that $\sqrt{n}l^{|\beta,\eta\rangle}_{\mathrm{inh}}(r) \gg 1$ is not satisfied for 
a range of relevant $n$ in regions of appreciable overlap between $W_{|\beta,\eta\rangle}(\alpha)$ and $W_{|n\rangle}(\alpha)$.
Specfically, for the amplitude-squeezed state we find $\sqrt{n}l^{|\beta,\eta\rangle}_{\mathrm{inh}}(r) > 1$ only for $n\gg100$, which explains the poor resemblance between the distributions 
$\tilde{P}_n$ and $P_n$. In contrast, for the phase-squeezed state we find $\sqrt{n}l^{|\beta,\eta\rangle}_{\mathrm{inh}}(r) \gg 1$ for $n \gg 25$, which is consistent with the failure of $\tilde{P}_n$ to reproduce the 
oscillatory structure of $P_n$ for $n \lesssim 25$.

\section{Application to Bose-Hubbard model\label{sec:BoseHubbard}}
In the previous sections we have considered illustrative examples with analytically known Wigner functions to justify the criteria for the validity of $\tilde{P}_n$. 
Here, we consider a numerical example -- the Bose-Hubbard model \cite{Gersch1963,Milburn1997} -- which demonstrates how a calculation of $\tilde{P}_n$ can enable physical insight in a non-trivial model, 
whilst the validity of $\tilde{P}_n$ can also be readily justified by examination of a numerically reconstructed single-mode Wigner function. 

The Bose-Hubbard model is in general not analytically tractable, and simple numerical methods such as exact diagonalization or solution of the Schr\"{o}dinger equation in a truncated Fock basis are 
generically only possible for a limited number of particles and/or sites. Hence, phase-space methods such as TWA have the potential to provide valuable insight into the dynamics of the system. 
For simplicity, we consider a two-site (two-mode) Bose-Hubbard model described by the Hamiltonian
\begin{equation}
 \hat{H} =  - \hbar\Omega\left(\hat{a}^{\dagger}_2\hat{a}_1 + \hat{a}^{\dagger}_1\hat{a}_2\right) +  \frac{\hbar U}{2} \sum_{i=1,2}  \hat{a}^{\dagger}_i\hat{a}^{\dagger}_i\hat{a}_i\hat{a}_i  , \label{eqn:H_BoseHubbard}
\end{equation}
where $U$ characterises the on-site self-interaction, $\Omega$ the tunneling strength between sites and $\hat{a}_i$ ($\hat{a}^{\dagger}_i$) is the usual bosonic
annihilation (creation) operator for sites $i=1,2$.  Due to the quartic nature of the interaction term, the evolution 
equation for the Wigner distribution will require truncation of third-order derivative terms. The impact of such truncation error is well understood in this 
context, with known signatures such as the inability of TWA to replicate revivals in population oscillations between modes \cite{OlsenRevivals2011}. We point out that we consider only the two-mode model 
in this instance so that truncation error can be monitored rigorously (via comparison to solution of the Schr\"{o}dinger equation in a truncated Fock basis). 
In general, one could apply the same procedure to a system with an arbitrary number of modes.

\begin{figure*}[tbp]
\includegraphics[width=18cm]{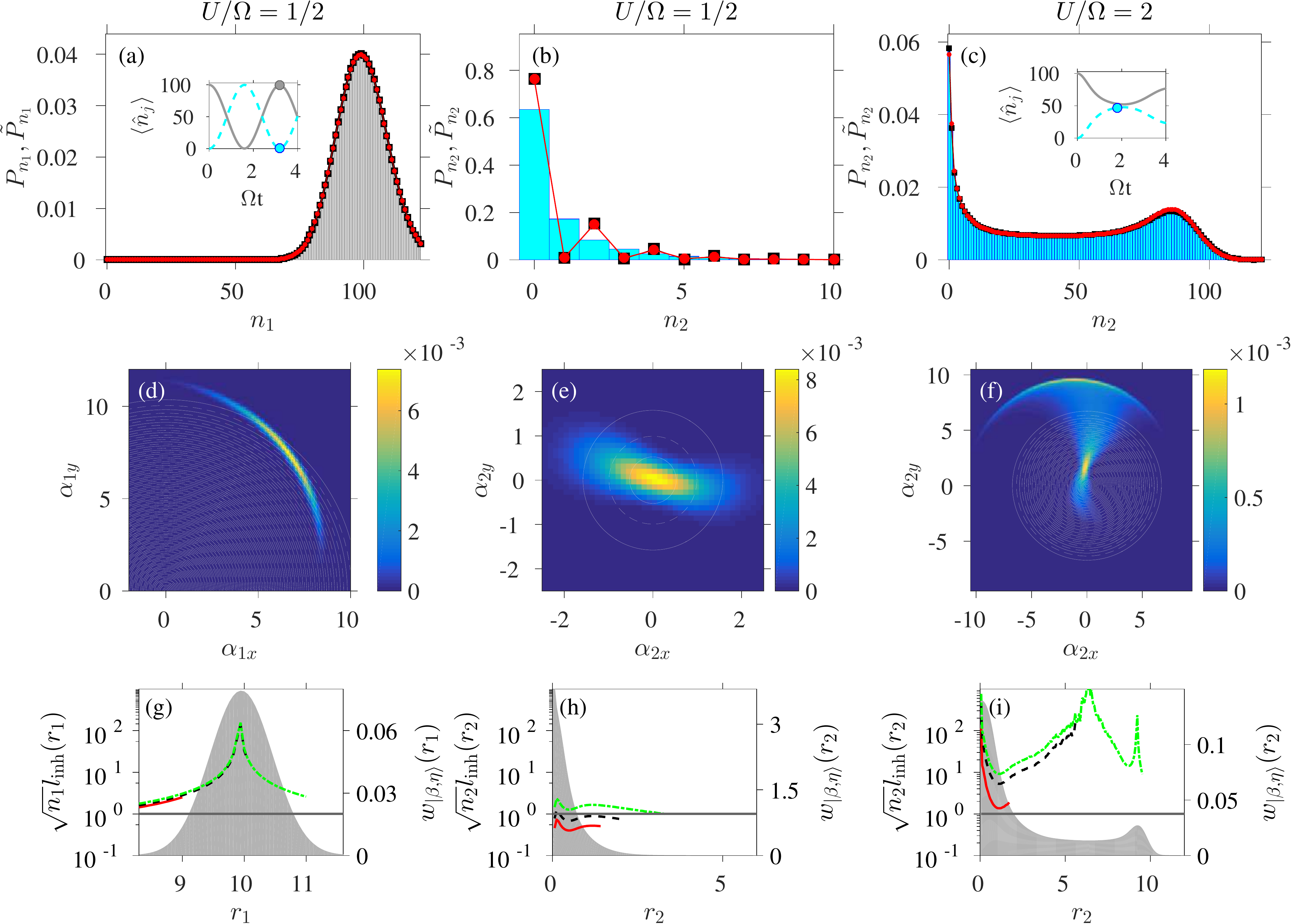}
\caption{(color online) Single-mode number distributions, (a) $\tilde{P}_{n_1}$ ($P_{n_1}$) and (b) $\tilde{P}_{n_2}$ ($P_{n_2}$) for $U/\Omega = 1/2$, and (c) $\tilde{P}_{n_2}$ ($P_{n_2}$) 
for $U/\Omega = 2$. Histograms indicate respective $\tilde{P}_{n_i}$, compared to $P_{n_i}$ calculated from: (i) exact diagonalization in truncated Fock basis (red circles) and (ii) TWA trajectories 
according to Eq.~\ref{eqn:Pn_WigTrick} [black squares, partially obscured by (i)]. Data points for $P_{n_i}$ are connected for visual clarity.
The insets of (a) and (c) shows the population evolution of the individual wells $\langle\hat{n}_1\rangle$ (grey solid line) and $\langle\hat{n}_2\rangle$ (cyan dashed line) for the respective cases, with 
the markers indicating the time at which the number distributions are evaluated. In plots (d)-(f) we show the single-mode integrated Wigner functions $W^{\mathrm{int}}_{|\psi\rangle}(\alpha_i)$ 
corresponding to the modes in (a)-(c). To better illustrate the relative radial width of these functions we overlay contours of single-mode Fock state Wigner functions $W_{|n_i\rangle}(\alpha_i)$ 
[minima (dashed lines) and maxima (solid lines) contours] with $n_i$ typical of the relevant range of values in the distribution: (d) $n_1 = 110$, (e) $n_2 = 3$ and (f) $n_2 = 47$. In (g)-(i) we plot the 
parameter $\sqrt{n_i}l^{|\psi\rangle}_{\mathrm{inh}}(r_i)$ for: (g) $n_1=80,100,120$, (h) $n_2=1,3,10$ and (i) $n_2=2,30,90$ (red solid, black dashed and green dot-dashed lines respectively on all plots). 
Comparison to $\sqrt{n_i}l^{|\psi\rangle}_{\mathrm{inh}}(r_i) = 1$ is indicated (dark grey line). 
To assist in identifying the region of overlap we terminate the lines at $r_i = \sqrt{n_i+1}$ to indicate the region containing $W_{|n_i\rangle}(r_i)$, whilst underlaying the 
radial distribution $w_{|\psi\rangle}(r_i)$ (shaded grey region) for comparison.
}
\label{fig:Pn_BoseHubbard} 
\end{figure*}

In Fig.~\ref{fig:Pn_BoseHubbard}, we compare the calculated single-mode distributions $P_{n_i}$ and $\tilde{P}_{n_i}$ \footnote{As previously noted in Sec.~\ref{sec:FormalDeriv}, 
one could also calculate the joint-probabilities $P_{n_i,n_j}$ and $\tilde{P}_{n_i,n_j}$ for this system, however, for simplicity of illustration we focus on the marginal distributions.} 
for a system initialized with all atoms in one mode ($i=1$), characterised by a coherent state of mean population $\langle \hat{n}_1 \rangle = 100$, and with: (a)-(b) $U/\Omega = 1/4$ and (c) $U/\Omega = 1$. 
We compare the binned distribution $\tilde{P}_{n_i}$ to $P_{n_i}$ calculated from: (i) exact diagonalization using a truncated Fock basis, and (ii) the TWA evolution and Eq.~(\ref{eqn:Pn_WigTrick}). 
Comparison between (i) and (ii) allows us to eliminate truncation error as the source of any potential difference between $\tilde{P}_{n_i}$ and $P_{n_i}$. 

We justify the validity (or invalidity) of $\tilde{P}_{n_i}$ by constructing the relevant single-mode integrated Wigner functions, 
$W^{\mathrm{int}}_{|\psi\rangle}(\alpha_i) \equiv \int d^2\alpha_j W_{|\psi\rangle}(\alpha_i,\alpha_j)$ where $W_{|\psi\rangle}(\alpha_i,\alpha_j)$ is the two-mode Wigner function 
of the state $|\psi\rangle$, from the TWA trajectories. From these, one may numerically construct the relevant radial distributions $w_{|\psi\rangle}(r_i)$ and associated inhomogeneity length scales 
$l^{|\psi\rangle}_{\mathrm{inh}}(r_i)$. It is then straightforward to quantitatively apply the criteria $\sqrt{n_i}l^{|\psi\rangle}_{\mathrm{inh}}(r_i) \gg 1$ for $r_i$ in the region of 
overlap of $w_{|\psi\rangle}(r_i)$ and $W_{|n_i\rangle}(r_i)$, for a specific $n_i$.

Following this procedure, we see that for the case of $U/\Omega = 1/2$ [at the evolution times indicated in the inset of Fig.~\ref{fig:Pn_BoseHubbard}~(a)] that $\tilde{P}_{n_ 1}$ is justified, as 
$\sqrt{n_1}l^{|\psi\rangle}_{\mathrm{inh}}(r_1) \gg 1$ for all $r_1$ in the region of overlap of $w_{|\psi\rangle}(r_1)$ and $W_{|n_1\rangle}(r_1)$ for the relevant $n_1\gtrsim 60$. In contrast, $\tilde{P}_{n_2}$ 
is a poor approximation to $P_{n_2}$ due to the positioning of $W^{\mathrm{int}}_{|\psi\rangle}(\alpha_2)$ at the origin and its relatively small radial width. We find that 
$\sqrt{n_2}l^{|\psi\rangle}_{\mathrm{inh}}(r_2) \gg 1$ is only satisfied for $n_2 \gg 10$, which is outside the scope of relevant $n_2$ for the distribution $P_{n_2}$.
By increasing the nonlinearity to $U/\Omega=2$, we find our method is also able to capture more complex number distributions, such as the twin-peaked structure seen in 
Fig.~\ref{fig:Pn_BoseHubbard}~(c). The broad structure of the underlying single-mode integrated Wigner function $W^{\mathrm{int}}_{|\psi\rangle}(\alpha_2)$, plotted in Fig.~\ref{fig:Pn_BoseHubbard}~(f) 
and quantified in Fig.~\ref{fig:Pn_BoseHubbard}~(h), justifies the use of $\tilde{P}_{n_2}$ for the overwhelming bulk of the distribution (negligibly small discrepancies are found for $n_2 \lesssim 3$).

\section{Conclusion}

In summary, 
we have examined under which conditions the binned number distribution from individual (truncated) Wigner trajectories, $\tilde{P}_n$, can replicate closely the 
true particle number distribution $P_n$. 
The sufficient requirement for this is that the Wigner function $W_{|\psi\rangle}(\alpha)$ of the state $|\psi\rangle$ varies sufficiently smoothly 
on the characteristic length scale of oscillations in the Wigner function $W_{|n\rangle}(\alpha)$ of the Fock state $|n\rangle$, defined quantitatively by the 
condition $l^{|\psi\rangle}_{\mathrm{inh}} \gg 1/\sqrt{n}$ in the region of overlap of the two Wigner distributions. 
This is, of course, in addition to the constraint 
that only positive Wigner functions $W_{|\psi\rangle}(\alpha)$ are being considered, which is the case in the truncated Wigner approximation or in model Hamiltonians that depend 
no-higher-than quadratically on creation or annihilation operators.

We have provided a rigorous operational definition of this seemingly heuristic binning procedure as one that corresponds to approximating the Wigner function of the Fock state 
(which appears in the definition of $P_n$ via an overlap integral with the Wigner function $W_{|\psi\rangle}(\alpha)$ of the state of interest) as a boxcar function in phase space.
For states localized around the phase-space origin (\textit{e.g.}, a thermal state), the requirement of smoothness of the Wigner function is satisfied by a broad distribution, 
having a characteristic width much larger than unity. In this case, the large width of the distribution is equivalent to having large mode occupation number. On the other hand, 
for states that have large coherent displacement $\beta$ (such as coherent and squeezed coherent states with $|\beta|\gg1$), one can tolerate a relatively narrow Wigner function, 
for $\tilde{P}_n\simeq P_n$, as long as its width remains much larger than $1/|\beta|$, which is the characteristic period 
of oscillations in $W_{|n\rangle}(\alpha)$ for the most relevant values of 
$n$ ($\sim|\beta|^2$). This condition is satisfied for coherent states and weakly squeezed states, but will break down for highly squeezed states when the width of the respective 
Wigner function $W_{|\beta,\eta \rangle}(\alpha)$ in the narrow dimension becomes comparable to $1/|\beta|$, even though the mode occupation for such states can be very high. 

Although we have considered only a small subset of states with analytically explicit Wigner functions in this article to illustrate our arguments, in Sec.~\ref{sec:BoseHubbard} we 
have also shown that our criteria under which the binned distribution $\tilde{P}_n$ can closely approximate $P_n$ is simply applicable to other, less trivial, states, such as 
those in the two-site Bose-Hubbard model. From a practical point of 
view, in the truncated Winger formalism the numerical reconstruction of an \textit{a priori} unknown single-mode Wigner function from many stochastic trajectories is relatively trivial 
and allows one to extract the characteristic length scale of the quasidistribution and thus, according to our criterion, accept
or reject the approximation $\tilde{P}_n$ with no knowledge of the exact $P_n$.


\begin{acknowledgments}
 R.~J.~L-S. acknowledges fruitful discussions with J.~F.~Corney, M.~J. Davis, S.~A.~Haine, S.~S.~Szigeti, and in particular M.~E.~Lewis. M.~K.~O. and K.~V.~K. acknowledge support by the Australia Research 
 Council Future Fellowships, grant Nos. FT100100515 and FT100100285, respectively.
\end{acknowledgments}


%

\end{document}